\documentclass[twocolumn,amsmath,amssymb,prl,showpacs,superscriptaddress]{revtex4}
\usepackage{graphicx}
\usepackage{sidecap}
\usepackage{dcolumn}
\usepackage{bm}

\begin{document}
\bibliographystyle{aip}
\title{Thermal Rectification in Graded Materials}
\author{Jiao Wang}
\affiliation{Department of Physics and Institute of Theoretical
Physics and Astrophysics, Xiamen University, Xiamen 361005, China}
\author{Emmanuel Pereira}
\affiliation{Departamento de F\'isica-ICEx, UFMG, CP 702, 30.161-970
Belo Horizonte MG, Brazil}
\author{Giulio Casati}
\email{giulio.casati@uninsubria.it}
\affiliation{ Center for Nonlinear and Complex Systems,
Universit\`a degli Studi dell'Insubria, Via Valleggio 11, 22100
Como, Italy}
\affiliation{Consorzio Nazionale Interuniversitario per le
Scienze Fisiche della Materia, e CNR-INFM}
\affiliation{Istituto Nazionale di Fisica Nucleare, Sezione di Milano,
via Celoria 16, 20133 Milano, Italy}
\date{\today}

\begin{abstract}
In order to identify the basic conditions for thermal rectification
we investigate a simple model with non-uniform, graded mass distribution.
The existence of thermal rectification is theoretically predicted
and numerically confirmed, suggesting that thermal rectification
is a typical occurrence in graded systems, which are likely to be
natural candidates for the actual fabrication of thermal diodes.
In view of practical implications, the dependence of rectification
on the asymmetry and system's size is studied.
\end{abstract}
\maketitle

The study of the underlying dynamical mechanisms which determine
the macroscopic properties of heat conduction has opened the
fascinating possibility to control the heat current. In particular
a model of a thermal rectifier has been proposed \cite{Casati2}
and since then, the phenomenon of thermal rectification has been
intensively investigated \cite{BLi1, BHu1, BLi2, BLi3, HBLi, Moha,
Prapid, P2011} in order to analyze and improve the rectification
effect, including experimental realizations \cite{Chang}.

However, as correctly pointed  out in Ref. \cite{BHu1}, most recurrent
proposals of a thermal diode, based on the sequential coupling of
two or three segments with different anharmonic potentials, are
difficult to be experimentally implemented and the rectification
power typically decays to zero with increasing the system size.
In addition, most investigations so far have been based on numerical
simulations and a much better theoretical understanding is highly
desirable both for fundamental reasons as well as for obtaining
useful hints for the actual realization of devices with
satisfactory rectification power.

Along these lines, graded materials  are attracting more and more
interest: Papers with numerical \cite{BLi3, HBLi, Moha}, analytical
\cite{Prapid, P2011} and even experimental \cite{Chang} studies
have appeared recently in the literature.


The present paper addresses the fundamental dynamical
mechanisms that lead to rectification. Our strategy is to consider
a simple model that contains the minimal ingredients we theoretically
judge to be necessary to rectification, and compare the numerical
results with the theoretical predictions. As the features of our model
are also shared by more realistic models such as anharmonic chains of
oscillators, we conjecture that the obtained results may have practical
implications as well. Our study allows us to understand the basic
ingredients behind rectification, to describe non-trivial and
important properties of the heat flow, and to show that rectification
in graded materials could be a ubiquitous phenomenon.

\begin{figure}
\includegraphics[width=7.5cm]{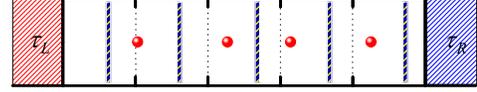}\caption{(Color online) The
schematic  plot of our model. Dotted lines divide elementary unit
cells. In each cell there is a bar which is subjected to  elastic
collisions with both cell boundaries and with neighboring bullets.
The first (last) cell is coupled to a heat bath at temperature
$\tau_L$ ($\tau_R$).}
\label{model}
\end{figure}

We consider a chain \cite{LCWP} of elastically colliding particles
of two kinds  referred to, in the following, as ``bars'' and
``bullets'', respectively. (See Fig. 1.) Each bar is confined
inside a cell of unit length; that is, besides elastic collisions
with its neighboring bullet(s), it is also subject to elastic
collisions with the two boundaries of the cell. A bullet only
undergoes collisions with its two neighboring bars. Apart from
collisions, all particles move freely. We denote the masses
(velocities) of bars by $M_l$ ($v_l$), $l=1,\cdots,Z$, and those
of bullets by $m_k$ ($u_k$), $k=1,\cdots,Z-1$, respectively, with
$Z$ being the total number of cells. Two statistical thermal baths
with different temperatures $\tau_L$ and $\tau_R$ are put into
contact with the two ends of the system: When the first (last) bar
collides with the left (right) side of the first (last) cell, it
is injected back with a new velocity determined by the
distribution \cite{Heatbath}
\begin{equation}
P_{L,R}(v) = \frac{|v|M_{1,Z}}{k_B \tau_{L,R}}\exp\left(
- \frac{v^{2} M_{1,Z}}{2 k_B \tau_{L,R}}\right).
\end{equation}
The Boltzmann constant $k_B$ is set to be unity. In our molecular
dynamics simulations, after the system reaches the steady state,
we can compute the local temperature of each cell and the steady
heat flux across the system. The local temperature of the $l$th
cell is defined as the kinetic temperature of the bar, i.e.,
$\tau_l\equiv \langle M_l v_l^2/k_B \rangle$, where $\langle \cdot
\rangle$ stands for the time average. The steady heat flux is
measured as the time average of the energy exchanged in unit time
between the left heat bath (the last bar) and the first bar (the
right heat bath), or that between any two neighboring particles.
We have verified that the so-measured heat flux is the same as the
local heat flux carried by each particle, which is defined as
$\langle M_l v_l^3/2 \rangle$ ($\langle m_l u_l^3/2 \rangle$) for
the $l$th bar (bullet).

In order to investigate rectification, we consider the mass graded
version of the above chain.  Precisely, we will focus on a system
with mass graded bullets $m_{1} < m_{2}< \cdots <m_{Z-1}$ and
identical bars, $M_1=\cdots= M_Z=1$. In order to avoid boundary
effects due to coupling with the heat baths, the rectification
effects are measured numerically only over the $N$ central
cells, with the first (last) $Z_L$ ($Z_R$) not being considered.
Hence the effective system size is $N=Z-Z_L-Z_R$.

The starting point of our theoretical approach is the expression
for the local heat flow inferred from the homogeneous case.  We
recall that the Fourier law of thermal conduction has been shown
numerically to hold in the homogeneous version (i.e., the bullets
also have the same mass $m_l=m$) of our model \cite{LCWP}. We also
recall that homogeneous lattice models of oscillators with, e.g.,
quartic anharmonic on-site potentials, have been studied by
different analytical methods \cite{Kup, Spohn, AK, LiLi} and shown
to follow the Fourier law as well. It implies that for these
systems, in the steady state the local heat current at position
$x$ is proportional to the temperature gradient $ \mathcal{F}(x) =
-\mathcal{K}(x)\nabla T(x)$, where the heat conductivity
$\mathcal{K}(x)$ is a function of the local temperature $T(x)$
(and of other parameters). Precisely, for the homogeneous models
of anharmonic oscillators, such a local expression reads
\begin{equation}
\mathcal{F} =
 \left( T_{j}-T_{j+1} \right) /{\mathcal{C}\bar{T}_{j}^{\beta}},
\end{equation}
where $T_j$ is the local temperature at the $j$th site and
$\bar{T}_{j}^{\beta}=  (T_{j}^ {\beta}+T_{j+1}^{\beta})/2$ with
$\beta \approx 2$. For the theoretical study of  rectification in
our inhomogeneous chain, we start from the inhomogeneous version
of such equation, i.e., we write the heat flow from cell $j$ to
$j+1$ as
\begin{equation}
\mathcal{F}_{j,j+1} = -\mathcal{K}_{j}(\nabla T)_{j} =
\frac{\left(\mathcal{C}_{j}T_{j}^{\alpha}+
\mathcal{C}_{j+1}T_{j+1}^{\alpha}\right)}{2} \left( T_{j}-T_{j+1}
\right) .\label{fluxo}
\end{equation}
Note that the homogeneous version of our model obeys the Fourier
law with a thermal  conductivity that scales as $T^{1/2}$ (see the
following), in contrast to the $1/T^{2}$ behavior of the above
anharmonic oscillators models.

For the heat flow in the steady state we have
\begin{equation}
\mathcal{F}_{1,2} = \mathcal{F}_{2,3} = \cdots =
\mathcal{F}_{N-1,N}  \equiv \mathcal{F} \label{corrente}.
\end{equation}
Hence, by summing up the $(N-1)$ equations in (\ref{fluxo}) for
$j=1,2,  \cdots, N-1$, we get $\mathcal{F} =
\mathcal{K}(T_{1}-T_{N})/(N-1)$, where
\begin{eqnarray}
\mathcal{K} =\frac{(N-1)}{2} \left[\sum_{j=1}^{N-1}\frac{1}{\mathcal{C}_{j}T_{j}^{\alpha}+
\mathcal{C}_{j+1}T_{j+1}^{\alpha}}\right]^{-1}.
\label{condutividade}
\end{eqnarray}
Then the Fourier law follows in the case where the thermal
conductivity $\mathcal{K}$ remains finite when $N\rightarrow\infty$.

\begin{figure}
\hskip-0.4cm
\includegraphics[width=8.4cm,height=6cm]{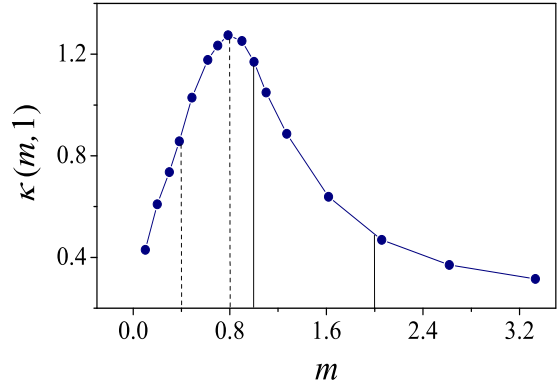}\caption{(Color
online) The numerically computed heat conductivity for the homogeneous
chain of size $N = 45$ at temperature $T=1$ (see Ref. \cite{LCWP}).
The bars have unit mass and the bullets have the same mass $m$. The
two solid (dotted) lines indicate the bullet mass region used for
the simulations in Fig. 3 (b) [Fig. 3(c)].}\label{fig2}
\end{figure}

From Eqs. (\ref{fluxo}) and (\ref{corrente}), it follows that
\begin{eqnarray}
&&(T_{1}-T_{2})(\mathcal{C}_{1}T_{1}^{\alpha}+\mathcal{C}_{2}T_{2}^{\alpha})
= (T_{2}-T_{3})(\mathcal{C}_{2}T_{2}^{\alpha}+\mathcal{C}_{3}T_{3}^{\alpha})
\nonumber \\
&& = \cdots = (T_{N-1}-T_{N})(\mathcal{C}_{N-1}T_{N-1}^{\alpha}+
\mathcal{C}_{N}T_{N}^{\alpha}) \label{tibau};
\end{eqnarray}
Thus, given the temperatures of the first and the last cell, i.e.,
$T_{1}$ and $T_{N}$, by  using the equations above we may
determine the inner temperatures $T_{2}$, $T_{3}$, $\cdots$,
$T_{N-1}$. As it may be a hard problem to obtain the analytical
solution of these equations, we may turn to numerical
computations, or else, as in Refs. \cite{Prapid,P2011}, we may
assume  small temperature gradients in order to simplify the
computations. In this latter regime, i.e., for $T_{1} = T +
a_{1}\epsilon$, $T_{N} = T + a_{N}\epsilon$, $\epsilon$ small, we
have $T_{k} = T + a_{k}\epsilon + \mathcal{O}(\epsilon^{2})$,
where $a_{k}$ have to be determined [we carry out the computations
only up to $\mathcal{O}(\epsilon)$]. Algebraic manipulations give
us $a_{k} = a_{1} + (a_{N}-a_{1}) \tilde{\mathcal{C}}(k)/
\tilde{\mathcal{C}}(N)$, for $k=2, \cdots, N-1$, where $ \tilde{
\mathcal{C}}(k)\equiv \left[(\mathcal{C}_{1}+\mathcal{C}_{2})^{-1}
+ (\mathcal{C}_{2}+ \mathcal{C}_{3})^{-1} + \cdots +
(\mathcal{C}_{k-1} + \mathcal{C}_{k})^{-1}\right]$. The thermal
conductivity defined in Eq. (\ref{condutividade}) is therefore
\begin{eqnarray}
\mathcal{K} &=& \frac{(N-1) T^{\alpha}}{2} \cdot \left[ \sum_{j=1}^{N-1}
  \left\{\frac{1} {(\mathcal{C}_{j}+ \mathcal{C}_{j+1})} \right.\right. \nonumber \\
&& - \left.\left.\frac{\alpha\epsilon}{T}\frac{(a_{j}\mathcal{C}_{j}+a_{j+1}
\mathcal{C}_{j+1})}{(\mathcal{C}_{j}+\mathcal{C}_{j+1})^{2}}\right\}
\right]^{-1}.
\end{eqnarray}

\begin{SCfigure*}
\includegraphics[width=12cm]{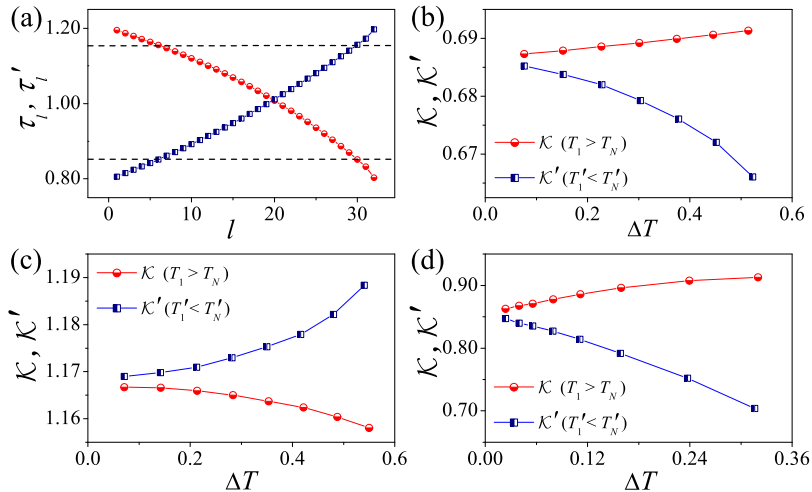}
\caption{(Color online) (a) Temperature profile of the graded
chain with bullets'  masses increasing in the interval (1,2) for
$\tau_L=1.2$ and $\tau_R=0.8$ (red dots), and inversely,
$\tau'_L=0.8$ and $\tau'_R=1.2$ (blue squares). The dashed lines
indicate the end temperatures $T_1$ ($T'_N$) and $T_N$ ($T'_1$) of
the central segment considered in our computations. (b)-(d) Heat
conductivity measured over the central segment of the graded chain
with (b) and (c) for bullets' masses increasing in the interval
(1,2) and  (0.4,0.8), respectively (all bars have unit mass), and
(d) for the graded chain where all the masses of bars and bullets
increase in the interval (1,2). See text for details.}
\label{fig3}
\end{SCfigure*}

Now we analyze the heat flow for the system with inverted thermal
baths.  In this case we denote the temperature of the $j$th cell
by $T'_j$ and assume that $T'_{1} = T_{N}$ and $T'_{N} = T_{1}$.
We also write $T'_{k} = T + a'_{k}\epsilon$ for $k = 2, 3, \cdots,
N-1$, and after the computations we get $ a'_{k} = a_{N} - \left(
a_{N}-a_{1} \right) \tilde{\mathcal{C}}(k)/
\tilde{\mathcal{C}}(N)$. Recalling that $a'_{1}=a_{N}$ and
$a'_{N}=a_{1}$, one obtains for the  thermal conductivity
$\mathcal{K}'$ of the ``inverted'' system, an expression similar
to $\mathcal{K}$ (with $a'_{j}$ replacing $a_{j}$). We then have:
\begin{eqnarray}
\lefteqn{\frac{1}{\mathcal{K}} - \frac{1}{\mathcal{K}'} =
\frac{2\alpha (T_{N}-T_{1})}{(N-1)T^{\alpha+1}\tilde{\mathcal{C}}(N)} } \label{difK}\\
&& \times \left\{ \frac{(\mathcal{C}_{1}-\mathcal{C}_{2})}{(\mathcal{C}_{1}+\mathcal{C}_{2})^{3}} +
\frac{(\mathcal{C}_{2}-\mathcal{C}_{3})}{(\mathcal{C}_{2}+\mathcal{C}_{3})^{3}} +
\cdots + \frac{(\mathcal{C}_{N-1}-\mathcal{C}_{N})}{(\mathcal{C}_{N-1}+
\mathcal{C}_{N})^{3}}\right\}. \nonumber
\end{eqnarray}

The homogeneous model, where all bars have a unit mass and all bullets
have mass $m$, has been shown \cite{LCWP} to follow the Fourier law
and the thermal conductivity $\kappa(m,T)$ at temperature $T=1$ is
as shown in Fig. 2. As the dynamics of our model can be essentially
described by a series of particle collisions, which is invariant
under a time rescaling $t\to t/\gamma$ (with $\gamma$ being a positive
constant), or equivalently a temperature rescaling $T\to \sqrt{\gamma}
T$, we have $\kappa(m,T) = \kappa(m,1)\sqrt T$. Therefore in the limit
of small temperature gradients, Eq. (\ref{fluxo}) leads to $\alpha =
1/2$. To conclude our theoretical analysis we still have to determine
the coefficients $\mathcal{C}_{j}$. The behavior of these coefficients
as function of $m$ can be deduced from Fig. 2 for the homogeneous chain.
Indeed let us also assume for our inhomogeneous system a small mass
gradient and determine such coefficients by properly relating them
to their homogeneous versions; i.e., $\mathcal{C}_{j} = [\kappa
(m_{j+Z_L-1},1)+\kappa (m_{j+Z_L},1)]/2$.

The qualitative picture which now emerges is quite clear: From
Eq. (\ref{difK}), in particular from the difference between
the local thermal conductivities, we predict, and will later
numerically confirm, the properties of rectification. First,
from Eq. (\ref{difK}) it immediately follows the existence of
rectification which is particularly clear when $\kappa(m_{j},1)$,
and so $\mathcal{C}_{j}$ has a monotonic behavior. It also follows
that rectification increases with the gradient of temperature.
Moreover, from Eq. (\ref{difK}), one can predict a larger flow
from a smaller to larger mass direction in the region where
$\kappa (m_{j},1)$ decreases with $m_{j}$, and the opposite
behavior in the region where $\kappa(m_{j},1)$ increases with
$m_{j}$. The fact that our approach allows to predict not only
rectification of the heat current but also its direction is quite
interesting. In this respect we recall that rectification in a
graded system with the heat current larger in the direction of
decreasing particle masses has been observed experimentally in
nanotubes with nonhomogeneous external mass loading \cite{Chang}
and numerically in the mass graded Fermi-Pasta-Ulam model
\cite{BLi3}. Instead, a case with the heat current larger in the
reversed direction (of increasing masses) has been found by
molecular dynamics simulations in mass graded ideal gases
\cite{HBLi} and some carbon nanotubes \cite{Moha}.

We now turn to the molecular dynamics studies of our system which
not only confirm  our theoretical predictions but also extend, in
fact, our analytical results to regimes beyond small gradients of
temperature and mass, where the rectification phenomena become
more relevant. To this end let us consider first a chain of $Z=32$
cells with the bullets' masses increasing in the range $(1,2)$ (i.e.
the interval between the two solid vertical lines in Fig. 2) with
$m_l=1+l/Z$ . After the steady state is reached we
calculate the time averaged temperature profile and the heat
current. Averages are taken over a number of collisions per
particle larger than  $2\times 10^{9}$. In Fig. 3(a) we show the
computed temperature profile for $\tau_L=1.2$ and $\tau_R=0.8$, as
well as for $\tau'_L=0.8$ and $\tau'_R=1.2$. To evaluate the heat
conductivity $\mathcal{K}$ and $\mathcal{K}'$ we neglect the first
(last) $Z_L$ ($Z_R$) cells in order to get rid of the boundary
effects. The values of $Z_L$ and $Z_R$ are determined by ensuring
that $T_1=T'_N$ and $T_N=T'_1$. (Note that $T_l=\tau_{l+Z_L}$ and
$T'_l=\tau'_{l+Z_L}$ for $l=1,\cdots,N$.) In this case we have
$Z_L=5$ and $Z_R=2$. [See Fig. 3(a).] The value of thermal
conductivity $\mathcal{K}=\mathcal{F}(N-1)/(T_1-T_N)$ is then
calculated over the central segment of $N=25$ cells with
numerically measured $\mathcal{F}$, $T_1$ and $T_N$.
$\mathcal{K}'$ is computed in the same way.

The dependence of the heat conductivity  $\mathcal{K}$ and
$\mathcal{K}'$ on $\Delta T$ for $T=1$ is given in Fig. 3(b). Here
$T_1=T'_N=T+\Delta T$ and $T_N=T'_1=T-\Delta T$. It is clearly seen
that a system with graded mass particles does present rectification,
and that rectification increases with the temperature gradient. In
addition, as in this case $C_j$ decreases monotonically (see the
region between two solid vertical lines in Fig. 2), according to
our theoretical analysis the value of $\mathcal{K}$ for $T_1>T_N$
should be larger than $\mathcal{K}'$ for $T'_1<T'_N$. This is
confirmed by the numerical results.

One advantage of our model is that, as shown in Fig. 2, it can
display regions where $C_j$ decreases with $j$ as well as regions
where $C_j$ increases with $j$ as, e.g., in the interval between
the two dashed vertical lines in Fig. 2. In the latter situation
we predict that the behavior of $\mathcal{K}$ and  $\mathcal{K}'$
will be opposite to the previous case, i.e., that $\mathcal{K}$
for $T_1>T_N$ will be smaller than $\mathcal{K}'$ for $T'_1<T'_N$.
This is confirmed by Fig. 3(c) where we present the numerical
results for a graded chain with the masses of bullets in the
interval $(0.4, 0,8)$ with $m_l=0.4(1+l/Z)$. All the computations
are done as before with $Z=32, Z_L=3, Z_R=4, N=25$ and $T=1$.

An obvious interesting question is how to enhance rectification
in our graded chain. As our theoretical analysis suggests, a
possible way to increase rectification is to increase the asymmetry
of the system. We have therefore analyzed the case where the masses
of the bars are also allowed to have a graded distribution. To be
precise, all the masses  are graded in between 1 and 2. Namely
$M_l=1+(l-0.5)/Z$ for $l=1, \cdots, Z$ and $m_l=1+l/Z$ for $l=1,
\cdots, Z-1$. Here we take $Z=32$, $Z_L=5$, $Z_R=1$, $N=26$ and $T=1$.
The results are depicted in Fig. 3(d) and show a significant increase
in rectification. We stress that the increase of rectification with
asymmetry is an important effect which may be of practical interest:
Similar mechanisms may be present in more realistic models due to
graded interparticle interactions and/or graded on-site potentials,
in addition to graded masses.

\begin{figure}
\hskip-0.4cm
\includegraphics[width=8.4cm,height=6cm]{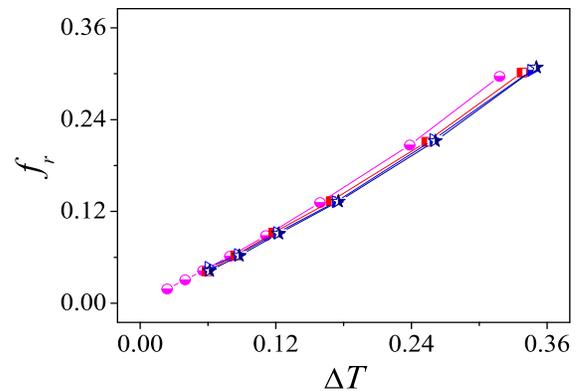}\caption{(Color
online) The rectification measured over the central segment of the
graded chain where all the masses of bars and bullets increase
linearly in the interval (1,2). The dots, squares, triangles and
stars represent the results for system size $Z=32, 64, 128$ and $256$,
respectively. }\label{fig4}
\end{figure}

Finally, we remark that for the case presented in Fig. 3(d) the
heat conductivity $\mathcal{K}$ and $\mathcal{K}'$ essentially
do not change with the system size. This  has been verified by
additional simulations with $Z=64, 128$ and $256$ (with $Z_L=8, 15$,
and $29$, $Z_R=2, 4$, and $8$, and $N=54, 109$, and $229$,
respectively). The rectification, defined as
$f_r\equiv (\mathcal{K}-\mathcal{K}')/\mathcal{K}'$, does not
change with the system size as clearly illustrated in Fig.
\ref{fig4}. The fact that rectification does not vanish with
increasing the  system size is an important property, which is 
absent in the known diode models given by the sequential coupling 
of different segments.

In summary, in this work, we have performed analytical and
numerical investigations of the heat flow in a simple model,
devoid of intricate interactions, in order to find results
of general validity. Our results demonstrate that thermal
rectification takes place in asymmetric systems with local
heat flow proportional to the temperature gradient and with
the local conductivity depending on temperature. This
conclusion is also supported by additional numerical
investigations \cite{unpublished}, along the lines of
Refs. \cite{LCWP} and \cite{Zhao}, of the energy diffusive behavior
and of correlation properties of energy fluctuations. Finally, 
it is interesting to notice that in our model the diffusive
behavior (typical of the Fourier law) is associated with the
thermal rectification phenomenon.

J.W. is supported by the NNSF (Grant No. 10975115) and SRFDP
(Grant No. 20100121110021) of China, E.P.  by CNPq (Brazil),
and G.C.  by the MIUR-PRIN 2008 and by Regione Lombardia.

\end{document}